\documentclass{sigchi}

\usepackage{balance}  
\usepackage{graphics} 
\usepackage{txfonts}
\usepackage{times}    
\usepackage[pdftex]{hyperref}
\usepackage{color}
\usepackage{textcomp}
\usepackage{booktabs}
\usepackage{ccicons}
\usepackage{todonotes}

\usepackage{paralist}

\makeatletter
\def\url@leostyle{%
  \@ifundefined{selectfont}{\def\UrlFont{\sf}}{\def\UrlFont{\small\bf\ttfamily}}}
\makeatother
\def\pprw{8.5in}
\def\pprh{11in}

\definecolor{linkColor}{RGB}{6,125,233}
\hypersetup{%
  pdftitle={Edvertisements: Adding Microlearning to Social News Feeds and Websites},
  pdfauthor={LaTeX},
  pdfkeywords={SIGCHI, proceedings, archival format},
  bookmarksnumbered,
  pdfstartview={FitH},
  colorlinks,
  citecolor=black,
  filecolor=black,
  linkcolor=black,
  urlcolor=linkColor,
  breaklinks=true,
}

\begin{document}

\title{Edvertisements: Adding Microlearning to \\ Social News Feeds and Websites}

\numberofauthors{3}
\author{%
  \alignauthor{Geza Kovacs\\
    \affaddr{Stanford University}\\
    \affaddr{Stanford, USA}\\
    \email{geza@cs.stanford.edu}}\\
}

\maketitle


\begin{figure}
\includegraphics[width=\columnwidth]{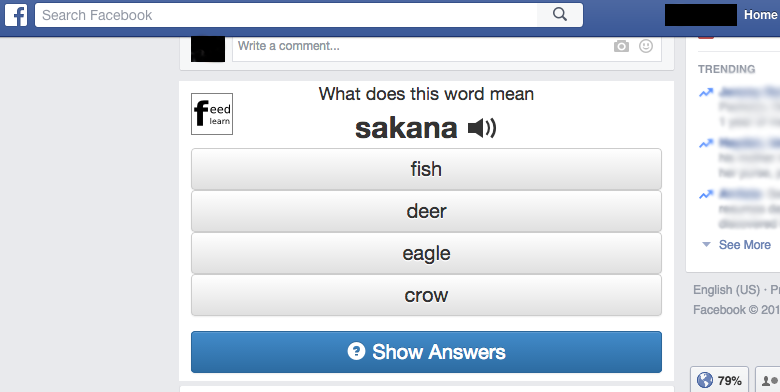}
\caption{Our extension can show interactive microlearning tasks (Edvertisements) in users' Facebook news feeds.}
\label{fig:feedlearn}
\end{figure}

\begin{figure}
\includegraphics[width=\columnwidth]{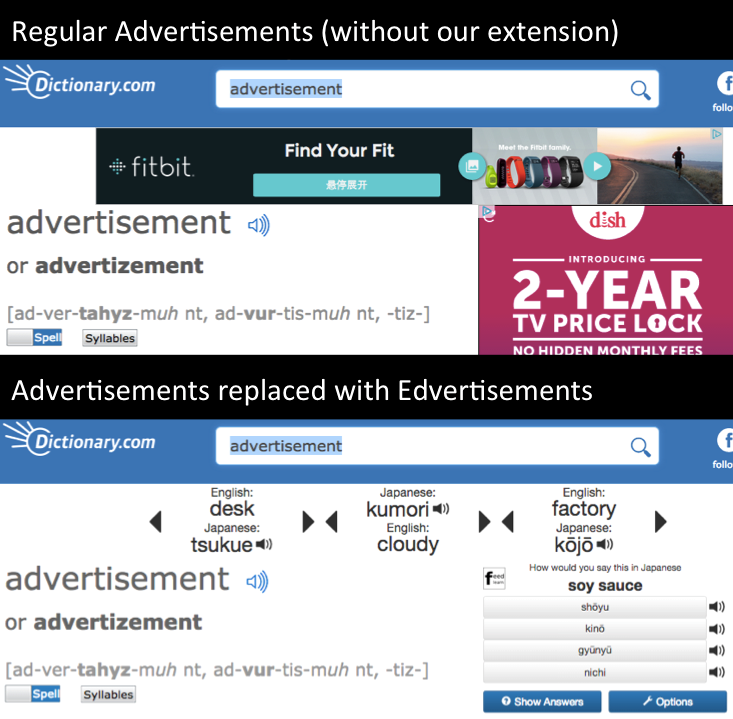}
\caption{Our extension can replace advertisements with interactive microlearning tasks (Edvertisements) on arbitrary websites.}
\label{fig:edvertisements}
\end{figure}

\begin{abstract}
Many long-term goals, such as learning a language, require people to regularly practice every day to achieve mastery. At the same time, people regularly surf the web and read social news feeds in their spare time. We have built a browser extension that teaches vocabulary to users in the context of Facebook feeds and arbitrary websites, by showing users interactive quizzes they can answer without leaving the website. On Facebook, the quizzes show up as part of the news feed, while on other sites, the quizzes appear where advertisements normally would.
In our user study, we examined the effectiveness of inserting microlearning tasks into social news feeds.
We compared vocabulary learning rates when we inserted interactive quizzes into feeds, versus inserting links that lead them to a website where they could do the quizzes. Our results suggest that users engage with and learn from our embedded quizzes, and engagement increases when the quizzes can be done directly within their feeds.
\end{abstract}

\keywords{microlearning; social feeds; facebook; advertisements; language learning}

\category{H.5.2.}{User Interfaces}{Graphical user interfaces (GUI)}{}{}


\section{Introduction}


Many people have long-term learning goals, such as wanting to learn a new language. However, they often fail to achieve these goals, citing the lack of time to study as a major reason \cite{micromandarin, kam2007localized, oxford1994language}.
Nevertheless, people do have spare time, as shown by their recreational web browsing and social network usage. 
American adults spend an average of 27 hours per month surfing the web \cite{nielsen2014}.
71\% of American adults use Facebook, and 63\% of these visit Facebook daily \cite{socialmediaupdate}. 90\% of American college students use Facebook, spending on average of 30 minutes on Facebook each day \cite{collegefacebook2}. 



In this paper, we present \textit{Edvertisements}, which help users learn during their spare time by showing them interactive microlearning tasks as they surf the web and read their Facebook feeds. We implemented a Chrome extension which shows Edvertisements in two ways:



\begin{compactitem}
\item On Facebook, the extension embeds Edvertisements directly into the feed alongside regular posts (see \autoref{fig:feedlearn}).
\item On other sites, the extension replaces web advertisements with Edvertisements (see \autoref{fig:edvertisements}). 
\end{compactitem}

\pagebreak

Our research questions are:

\begin{compactitem}
\item Do users engage with and learn from Edvertisements that we insert into their Facebook feeds?
\item Do users engage more with Edvertisements if they can complete the microlearning tasks without leaving their Facebook feeds (compared to external links)?
\end{compactitem}

In our user study, we examined engagement and vocabulary acquisition rates after embedding Edvertisements into users' Facebook feeds. We found that users interacted frequently with Edvertisements, improved their post-test results after a week, and engaged more readily with Edvertisements when they did not have to leave their social news feeds.

\section{Related Work}

\subsection{Microlearning}

Microlearning is the strategy of studying frequently in short intervals throughout the day \cite{gassler2004integrated}. Several mobile applications use microlearning to teach foreign language vocabulary and other materials \cite{microlearning, micromandarin}; however, these require the user to interrupt their routine to open a dedicated app for studying. 

Some systems try to solve this problem by embedding microlearning into other contexts. There are games in which users complete learning tasks while playing \cite{carriearcade}, video players which teach vocabulary while users watch foreign-language videos \cite{smartsubtitles}, screensavers that display facts while the screen is idle \cite{screensaver}, and chat clients that show vocabulary while the user is chatting \cite{cai2015wait}.


Compared to the learning contexts used by existing systems, we believe that web surfing and Facebook feeds are especially potent locations for embedding microlearning, because:

\begin{compactitem}
\item Unlike playing educational games or watching foreign-language videos, visiting Facebook is a daily habit for 45\% of American adults with an internet connection. \cite{socialmediaupdate} 
\item Web surfing and reading Facebook news feeds are recreational activities, so the embedded microlearning tasks will not interrupt users' work.
\item Users are already accustomed to a variety of rich content appearing in their Facebook feeds, such as videos, games, recommendations, and advertisements, so they should not find the added Edvertisements too distracting.
\end{compactitem}



\subsection{Using Social News Feeds to Trigger Desired Habits}


Many apps post on users' Facebook feeds to drive engagement. 
For example, Duolingo can share users' study progress, and Strava can share users' exercise history.
These posts aim to get users' friends to send them encouraging feedback, and to try the apps themselves.
However, these posts are often viewed as bragging about trivial accomplishments, and receive little attention \cite{socialsharing}. 

These posts are examples of \textit{triggers}, which are calls to action designed to help users form habits \cite{foggpersuasive}. Facebook app posts require the user to go to a different website to study, as Facebook's API does not allow apps to post interactive content. With Edvertisements, we lower the barrier to action by allowing the user to study without leaving the website.


\pagebreak

\subsection{Web Advertising and Ad-Blocking}

Although advertisements are an important revenue source for websites, surveys indicate that 77\% of users almost never click on ads, and 69\% express interest in skipping or blocking ads \cite{adblockinggames}.
16\% of US web users use ad blockers, which are browser extensions that prevent web ads from being displayed. Ad blocker usage is growing -- global ad blocking has more than tripled since 2013, and is posed to further grow as ad blockers for mobile devices gain traction \cite{costofadblocking}. 



In surveys, users of ad-blockers cite ``distracting animations and sounds'', and ``offensive/inappropriate ad content'' as their top reasons for blocking ads \cite{adblockinggames}. Even users who do not install ad blockers tend to avoid looking at ads, a phenomenon known as ``banner blindness.'' In fact, web surfers click on less than 0.5\% of advertisements -- a number which has been declining ever since banner ads were introduced in 1994 \cite{whypeopleavoidadvertising}. Edvertisements repurpose this space for microlearning.

\section{Edvertisements System}

Our system is a Chrome extension that can show users microlearning tasks -- in our case, vocabulary quizzes -- in users' Facebook feeds, or as they are browsing the web. Although we originally implemented the browser extension for Chrome, we have also ported it to Firefox, and our technique can be implemented on any browser that supports extensions (Chrome, Firefox, Edge, Safari, etc). Our system features a variety of microlearning tasks for learning vocabulary in multiple languages, but in this paper we will focus on learning Japanese vocabulary.



\subsection{Inserting Edvertisements into Facebook Feeds}

Our extension can insert Edvertisements into users' Facebook feeds as rectangular interactive quizzes mimicking the look of a regular feed item, as shown in \autoref{fig:feedlearn}. We chose to insert 1 microlearning task for every 10 normal feed items, to mimic the approximate frequency we observed sponsored content appearing in the feed.

\subsection{Replacing Web Advertisements with Edvertisements}

People spend considerable time on sites other than Facebook, so we also created a general mechanism for presenting microlearning tasks as users browse the web. Our extension detects web advertisements on pages, and replaces them with microlearning tasks.


We detect the presence of web advertisements using the same approach as ad blockers -- by checking the URL the element originates from, and comparing it against EasyList, a list of known URL patterns for advertisements maintained by Adblock Plus. When we detect an element that is an advertisement, we replace it with an Edvertisement of the same size.

Web advertisements follow standardized sizes, called the IAB (Interactive Advertising Bureau) Standard Ad Units.
We have created microlearning tasks which fit 2 of the common sizes -- 300x250 and 200x90 -- corresponding to regular-sized and small ads.  If a microlearning task in the appropriate size is not available, we pick a smaller one and scale and stack it to fit the available space. For example we can fill a banner ad (728x90) with 3 small Edvertisements, as shown in \autoref{fig:edvertisements}. 





\pagebreak

\subsection{Quiz Types}

\begin{figure}
\centering
\includegraphics[width=1.0\columnwidth]{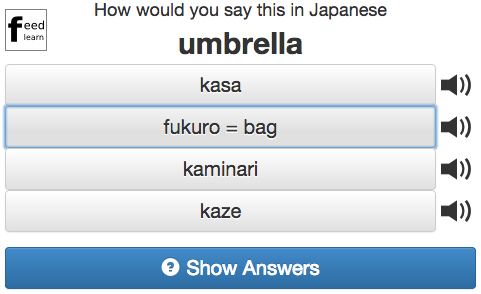}
\caption{One type of quiz presents a noun in English (umbrella), and asks the user to select the correct translation into Japanese (\textit{kasa}). The user has incorrectly selected \textit{fukuro}, so the user is shown its meaning (bag), and tries again.}
\label{fig:quiz2}
\end{figure}

\begin{figure}
\centering
\includegraphics[width=1.0\columnwidth]{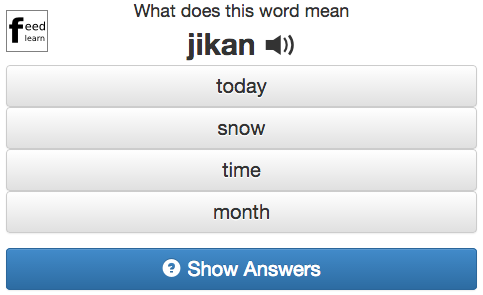}
\caption{Another type of quiz presents a noun in Japanese (\textit{jikan}), and asks the user to select its meaning (time).}
\label{fig:quiz1}
\end{figure}

One type of quiz presents a noun in English, and asks the user to select the corresponding Japanese word, as shown in \autoref{fig:quiz2}. To ensure that users learn word associations in both directions, we also have a second type of quiz, which shows the user a Japanese word and asks for the corresponding English translation, as shown in \autoref{fig:quiz1}.

We chose this multiple-choice quiz format because it tests the user's knowledge with a minimal amount of interaction -- the user simply clicks on a word to answer. Once the user answers a quiz correctly, a new quiz testing a different word is displayed. Thus, users can engage with an Edvertisement to continue study vocabulary for as long as they wish to.



 \subsection{Quiz Generation}

We obtained words and translations from the Nouns section of Wiktionary's 1000 Basic Japanese Words list. We excluded loanwords that users would easily recognize (\textit{pinku}=pink), and words that become homographs when romanized (\textit{hana}=flower or nose). We focus on nouns, because they are the most common type of word \cite{microlearning}. 

\pagebreak

\subsection{Spaced Repetition}

Spaced repetition algorithms schedule items for review to ensure long-term retention \cite{karpicke2011spaced}. We modified the Memreflex algorithm \cite{memreflex} to show the overdue word that appeared least recently in the feed (or to introduce a new word if there are no overdue words), instead of always showing the most overdue word. 
This ensures that users will continue to see different words as they are scrolling through their feeds, even if they are not always answering the in-feed quizzes.





\section{User Study}

We conducted a preliminary user study to see how frequently users would engage with Edvertisements, and compare the effectiveness of embedded interactive quizzes that can be completed without leaving the page, versus inserting static links as is done by today's advertisements and sponsored Facebook posts. 


In our study, we only inserted Edvertisements into Facebook feeds, and did not manipulate advertisements, since many users were already using ad-blockers. Furthermore, manipulating the Facebook feed enabled us to better control the frequency and size of inserted quizzes, compared to repurposing existing advertisement slots. 


\subsection{Participants}

We recruited 14 users (6 female, 8 male; ages 18-29 with median age 20) who had not previously studied Japanese but were interested in learning some basic vocabulary. They were voluntary participants recruited from online forums and Facebook groups related to Japanese culture. All of our participants self-reported that they were regular users of Facebook. 



\begin{figure}
\centering
\includegraphics[width=1.0\columnwidth]{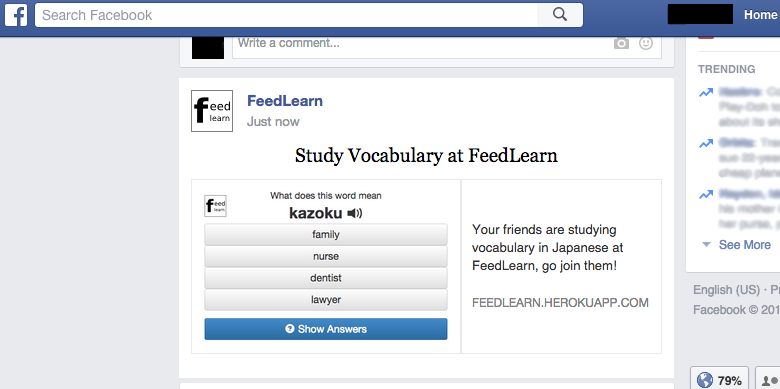}
\caption{The control condition in our user study inserted a link into users' Facebook feeds that led them to a site where they could do quizzes.}
\label{fig:control}
\end{figure}

\subsection{Materials}

We selected 50 basic Japanese words from Wiktionary's Basic Japanese Words list as the study material. We presented vocabulary words in romanized form instead of Japanese script, since our users could not read Japanese script.


\subsection{Conditions}

Users were assigned to one of two conditions:

\begin{compactitem}
\item Users in the \textit{in-feed quiz} condition had quizzes inserted directly into their feeds, as shown in \autoref{fig:feedlearn}.
\item Users in the \textit{link} condition were shown links to a site where they could do the quizzes, as shown in \autoref{fig:control}.
\end{compactitem}

Apart from the different items (quizzes/links) inserted into the feed, the questions and quiz interfaces were identical in both conditions. In both conditions, the items were inserted at a rate of 1 quiz/link per 10 feed items. 


\subsection{Procedure}

The study was conducted entirely online. First, users took a pre-test where they tried matching the 50 Japanese words to their 50 English definitions. Then they installed our Chrome extension and used it to study the 50 words for a week. After a week, we asked users to take the post-test, which had the same format as the pre-test.


\section{Results}

\subsection{Vocabulary Quiz Results}

\begin{figure}
\centering
\includegraphics[width=1.0\columnwidth]{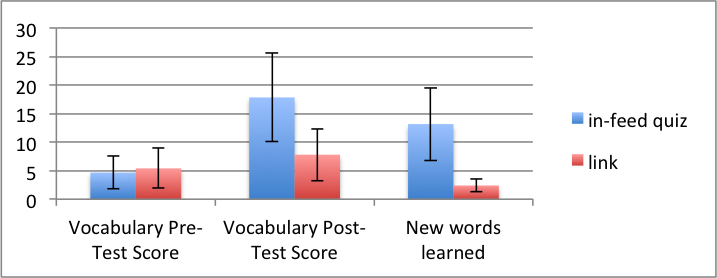}
\caption{Vocabulary test scores for the in-feed quiz and link conditions, with standard error bars.}
\label{fig:vocab-test-scores}
\end{figure}

\autoref{fig:vocab-test-scores} shows average vocabulary pre-test and post-test scores. On average, users in the in-feed quiz condition learned 13.2 new words, compared to 2.4 new words in the link condition. However, this difference was not statistically significant (t=1.38, p=0.20).

\subsection{Engagement With Edvertisements}

\begin{figure}
\centering
\includegraphics[width=1.0\columnwidth]{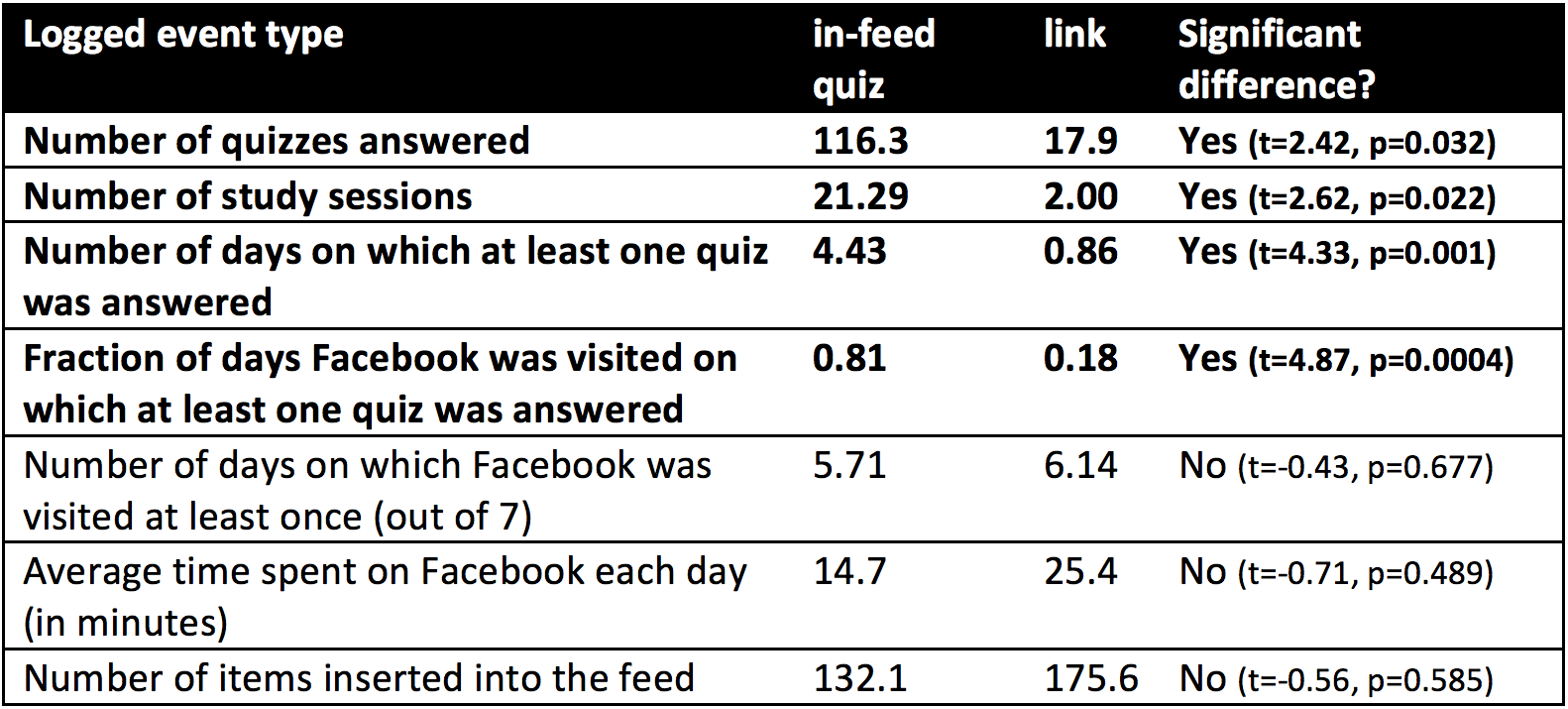}
\caption{Average number of events logged per user for the in-feed quiz and link conditions.}
\label{fig:event-logs}
\end{figure}

\autoref{fig:event-logs} shows the number of times users practiced answering quizzes. We also kept track of ``study sessions'', which we defined as the number of times the user clicked on the link to visit the website (in the link condition), or first engaged with an Edvertisement (in the in-feed quiz condition). 

We found that there was high engagement with in-feed Edvertisements -- on average, users answered 116 quizzes across 21 study sessions and answered a question 4.4 days of the 5.7 days they visited Facebook. Users in the in-feed quiz condition answered significantly more quizzes than the link condition, engaged in more study sessions, and studied on more days across the week. 

\pagebreak

\subsection{Qualitative Feedback}


Some users mentioned that they would prefer words to be explicitly introduced first before they start appearing in quizzes.  In addition, as we can see from the ratio of study sessions to feed insertions, even in the in-feed quiz condition, users only interact with 1/5 of quizzes they see. Hence, we need to ensure that seeing items reinforces memory, even if users do not interact with them. To address this issue, we later added new types of items to introduce new words and review old ones, such as the one shown in \autoref{fig:edvertisements}.

\section{Conclusion and Future Work}

Edvertisements are microlearning tasks that we show to users as they are surfing the web. We have built a browser extension that can insert Edvertisements into Facebook feeds, or replace web advertisements with Edvertisements.


In our user study, we inserted Edvertisements teaching vocabulary into users' Facebook feeds. We found that users engaged with and learned from Edvertisements, and that engagement rates were significantly higher when the quizzes could be done without leaving their Facebook feeds.

Our current implementation can insert Edvertisements into Facebook feeds, and can replaces web advertisements with them. There are other online contexts where we might show users Edvertisements -- for example, between Youtube videos, chapters of an e-book, or in their email. Although we have focused on microlearning, Edvertisements could also be used to remind people to do other small, beneficial tasks as they are idly surfing the web -- for example, encouraging people to  do a short exercise, or complete an item on their to-do list. 

Other directions for future work include making Edvertisements more personalized and contextually relevant, based on the user's browsing history and social networks. For example, if today is your Chinese friend's birthday and your browsing habits indicate that you are learning Chinese, we might show an Edvertisement in your Facebook feed teaching you how to wish him a happy birthday in Chinese. Or if you are reading an anti-vaccination webpage, an Edvertisement might teach you scientific facts about vaccines, showing the names of your friends who have also completed that Edvertisement. Or if many of your friends run, we might show you an Edvertisement about the health benefits of running, showing your friends' recent runs as part of the Edvertisement. 

Production of Edvertisements and monetization is another area of future work. One potential approach is sponsored Edvertisements -- for example, a local gym might sponsor an Edvertisement which teaches you a workout routine, hoping it will encourage you to visit their gym.


People spend hours surfing the web, but many fail to invest time towards learning and other forms of self-improvement. By putting interactive microlearning tasks right in peoples' social feeds and websites, we aim to help users learn in their spare time, one Edvertisement at a time. 

\section{Edvertisements Demo and Source Code}

Researchers interested in using or building on Edvertisements can visit \url{https://edvertisements.github.io/}

\pagebreak

\balance{}

\bibliographystyle{SIGCHI-Reference-Format}
\bibliography{edvertisements}

\end{document}